# Longitudinal spin current absorption in bilayers composed of ferromagnetic and highly-resistive non-magnetic layers

Sosuke Hori[1], Kohei Ueda[1,2,3,a)], Junichi Shiogai[1,3], and Jobu Matsuno[1,2,3]

[1]*Department of Physics, Graduate School of Science, Osaka University, Osaka 560-0043, Japan*

[2]*Center for Spintronics Research Network, Graduate School of Engineering Science, Osaka University, Osaka 560-8531, Japan*

[3]*Division of Spintronics Research Network, Institute for Open and Transdisciplinary Research Initiatives, Osaka University, Suita, Osaka 565-0871, Japan*

**ABSTRACT (< 150 words)**

Spin Hall magnetoresistance (SMR) is an intriguing spin-dependent transport phenomenon in bilayers consisting of non-magnetic and magnetic layers. Here, we report on the influence of longitudinal spin current absorption by the magnetic layer on SMR in bilayers composed of $Co_{20}Fe_{60}B_{20}$ (CoFeB) and epitaxial $SrIrO_3$, where $SrIrO_3$ is used as a highly-resistive spin current source. We observed a clear SMR signal and an enhancement in the SMR ratio with increasing CoFeB layer thickness, in qualitative agreement with an SMR model that incorporates the spin current absorption. The effective spin Hall angle is corrected from 0.07 to 0.12 with consideration of the spin current absorption, corresponding to a relative correction of ~71%. Our findings highlight the pronounced impact of the spin current absorption by the magnetic layer on the SMR mechanism when employing highly-resistive non-magnetic layer such as $SrIrO_3$, as well as other emerging quantum materials. (144 words)

[a)] kueda@phys.sci.osaka-u.ac.jp

## 1. Introduction

Non-magnet (NM) with strong spin-orbit coupling enables efficient spin current generation via the spin Hall effect (SHE)[1)]. The SHE is also found to give rise to an intriguing magnetoresistance, namely, spin Hall magnetoresistance (SMR)[2-13)]. Over the past decade, the SMR has served as a pivotal technique for investigating spin transport properties in bilayer systems consisting of NM and magnetic layers[2-13)]. The observation of SMR was first reported in a bilayer utilizing insulating ferrimagnet $Y_3Fe_5O_{12}$ (YIG) and Pt with a large SHE[2)]. In SMR, longitudinal resistance depends on the relative orientation between the spin current polarization $\sigma$ and the magnetization $M$. When $\sigma$ is perpendicular to $M$ ($M \perp \sigma$, so-called transverse spin current), the resistance is high due to absorption of the spin current by the magnetic layer since spin current acts on $M$ via interfacial spin mixing. When the $\sigma$ is parallel to the $M$ ($M$ // $\sigma$, longitudinal spin current), the spin current is reflected at the interface without absorption. Subsequently, the reflected spin current is converted back into a charge current by the inverse SHE, resulting in lower resistance as sketched in Fig. 1(a).

Following the extensive research on Pt/YIG systems, the SMR investigations have expanded to encompass bilayers consisting of metallic ferromagnet (FM) such as CoFeB alloy[5,7,8–10)] and Co[4,9)]. In contrast to the insulating YIG, the metallic FM layer also absorbs the longitudinal spin current as previously reported[5,7)]. Consequently, the reflected spin current is reduced at the interface, resulting in higher resistance compared to the case of insulating magnetic layer in the longitudinal spin current condition as illustrated in Fig. 1(b). Given this contribution, the SMR ratio $\Delta R_{\mathrm{SMR}}/R_0$ is expressed in Eq. (1) derived from the spin-drift-diffusion model[5,7)].

$$\frac{\Delta R_{\mathrm{SMR}}}{R_0} = -x_{\mathrm{NM}}\xi_{\mathrm{SH}}^2 \frac{\lambda_{\mathrm{NM}}}{t_{\mathrm{NM}}} \tanh^2\left(\frac{t_{\mathrm{NM}}}{2\lambda_{\mathrm{NM}}}\right)\left[\frac{1}{\coth(t_{\mathrm{NM}}/\lambda_{\mathrm{NM}})} - \frac{g_{\mathrm{FM}}}{1+g_{\mathrm{FM}}\coth(t_{\mathrm{NM}}/\lambda_{\mathrm{NM}})}\right] \quad (1)$$

$$g_{\mathrm{FM}} = \frac{(1-P^2)\rho_{\mathrm{NM}}\lambda_{\mathrm{NM}}}{\rho_{\mathrm{FM}}\lambda_{\mathrm{FM}}\coth(t_{\mathrm{FM}}/\lambda_{\mathrm{FM}})}$$

Here, $t_{\mathrm{NM}}$, $t_{\mathrm{FM}}$, $\rho_{\mathrm{NM}}$, $\rho_{\mathrm{FM}}$, $\lambda_{\mathrm{NM}}$, and $\lambda_{\mathrm{FM}}$ are the thicknesses, resistivities, and spin diffusion lengths of the NM and FM layers, respectively. $P$ denotes the spin polarization of the current in the FM layer. $x_{\mathrm{NM}}$ indicates the current fraction flowing through the NM layer defined as $\rho_{\mathrm{FM}}t_{\mathrm{NM}}$ / ($\rho_{\mathrm{FM}}t_{\mathrm{NM}} + \rho_{\mathrm{NM}}t_{\mathrm{FM}}$). $\xi_{\mathrm{SH}}$ is the effective spin Hall angle in the NM layer. The first term in square bracket in Eq. (1) represents the conventional SMR model used for insulating magnetic layer, where absorption of the longitudinal spin current by magnetic layer is prohibited ($g_{\mathrm{FM}} \sim 0$) [Fig. 1(a)]. The second term accounts for the extended model that incorporates the absorption of longitudinal spin current [Fig. 1(b)]. According to Eq. (1), without consideration of the second term, the value of $\xi_{\mathrm{SH}}$ can be underestimated in the presence of the absorption of longitudinal spin current, for instance, when $\rho_{\mathrm{NM}}$ is comparable to or larger than $\rho_{\mathrm{FM}}$. Despite the critical role of spin current absorption in SMR analysis, most of the studies have so far focused on bilayers composed of two metallic layers where the condition $\rho_{\mathrm{FM}} >> \rho_{\mathrm{NM}}$ is fulfilled.

Recently, a large SHE has been achieved in 5$d$ transition metal oxides such as $SrIrO_3$ (SIO)[14-21)], $IrO_2$[12,22–24)], and $WO_2$[25)]; our previous study has also shown the efficient spin current generation in a bilayer composed of CoFeB[18)] and epitaxial SIO. Moreover, recent advances have highlighted the efficient spin current generation from bulk and/or interface of many quantum materials such as topological semimetal[26,27], topological insulator[28–30], and two-dimensional material[31,32]. While these intriguing NM typically exhibit higher resistivity than often-used metallic FM layers such as CoFeB[5,7,8,10] and Co[4,9], the role of the longitudinal spin current absorption in SMR experiments in these systems remains an open question. Therefore, clarification of spin transport mechanisms in a bilayer composed of the highly-resistive NM materials provide a valuable contribution for further development of spintronics.

In this letter, we report on the effect of the absorption of longitudinal spin current by magnetic layer on SMR in CoFeB/SIO bilayers. Because SIO exhibits a large spin Hall angle and its resistivity of 570 μΩcm is larger than that of CoFeB (190 μΩcm) satisfying the condition of $\rho_{NM}$ being larger than $\rho_{FM}$[18], the CoFeB/SIO bilayer serves as an ideal model system for further elucidating the influence of the magnetic layer on SMR. A clear SMR signal is observed by measuring the angular dependence of the longitudinal resistance under a fixed external magnetic field in the *zy* plane. The SMR ratio is found to increase with increasing CoFeB layer thickness, which can be explained by the absorption of magnetic layer. The effective spin Hall angle has a relative correction of ~71% after the consideration based on the SMR model. These findings show that the magnetic layer exerts a significant influence on SMR in systems employing highly-resistive NM layers.

## 2. Sample preparation

Following our previous study[18], the SIO films were grown on $DyScO_3$ (DSO) substrate by a pulsed laser deposition from ceramic SIO target using a KrF excimer laser ($\lambda$ = 248 nm) at 5 Hz. Substrate temperature and oxygen pressure during the deposition were 650℃ and 25 Pa, respectively. The crystal structure of bulk SIO and DSO is an orthorhombic $GdFeO_3$-type perovskite with the pseudocubic lattice parameters of 0.3950 nm[33] and 0.3942 nm[34], respectively. The small lattice mismatch of −0.2% enables the epitaxial growth of SIO on the DSO(110) substrate with a compressive strain. The high-quality epitaxial SIO film was confirmed by x-ray diffraction (XRD) measurement; reciprocal lattice mapping reveals that in-plane lattice constant of the film is locked to the substrate, indicating a coherent growth of SIO film. The high-quality growth of $SrIrO_3$ is further supported by atomic force microscopy (AFM) measurements, showing atomically flat surfaces with a root mean square (RMS) roughness of 0.2 nm (see Appendix A1). This results in a bilayer with smooth interface roughness, providing a suitable platform for investigating the SMR. Additionally, a resistivity measurement shows weak temperature dependence and a room-temperature value ~600 μΩcm, which is consistent with previous reports on stoichiometric SIO films[21,42–44] (see Appendix A1). It has been reported that Ir cation deficiency in SIO, often induced by low oxygen partial pressure during deposition, significantly increases the resistivity, often exceeding 2 mΩcm at room temperature[21]. In contrast, the relatively low resistivity (~600 μΩcm) observed in our films suggests that Ir vacancies are negligible. These characterizations confirm that our SIO films fulfill the experimental requirement for a highly resistive spin current source, in contrast to typical 5*d* TMs.

To perform SMR measurements, bilayers were prepared by depositing CoFeB on SIO thin films. The overall structure of the samples is $TaO_x$(2.5)/CoFeB($t_{CoFeB}$ = 2 – 6)/SIO(21) shown in Fig. 2(a), where the numbers in parentheses are layer thickness in nm. Ta/CoFeB was deposited *ex situ* by rf magnetron sputtering. $TaO_x$ is obtained by atmospheric oxidation of the Ta layer and serves as a cap layer to protect CoFeB from oxidation. While the thickness of SIO was determined from the Laue fringes period by XRD measurement, the thickness of Ta and CoFeB was determined from the deposition rate measured in advance by x-ray reflectivity measurement. After deposition, the CoFeB/SIO bilayers were patterned into Hall-bar devices by photolithography and Ar ion milling. Pt(60)/Ta(5) contact pads were attached at the ends of devices for electrical measurement. The Hall bar has channel dimensions of $L$ = 30 μm and $w$ = 10 μm width, as shown in the inset of Fig. 2(b). The longitudinal resistances $R$ of CoFeB/SIO bilayers were measured by lock-in technique. The $x_{NM}$ was found to vary from 0.79 to 0.52 at $t_{CoFeB}$ from 2 nm to 6 nm in the CoFeB/SIO bilayers as shown in Fig. 2(b).

## 3. Results and discussion

The angular dependence of $R$ was measured to evaluate magnetoresistance while rotating the external magnetic field $B_{ext}$. When a charge current $I_c$ is applied along the $x$-direction as shown in Fig. 2(c), $\sigma$ induced by the SHE points in the $y$-direction. In this configuration, the $R$ is given as follows[4].

$$R = R_0 + \Delta R_{AMR} \sin^2\theta_M \cos^2\phi_M + \Delta R_{SMR} \sin^2\theta_M \sin^2\phi_M \quad (2)$$

where $\theta_M$ and $\phi_M$ are the polar and azimuthal angle of the magnetization $M$ of the CoFeB layer, respectively, $R_0$ is the resistance at $\theta_M$ = 0, and $\Delta R_{AMR}$ and $\Delta R_{SMR}$ is the magnetoresistance stemming from anisotropic magnetoresistance (AMR)[35] and SMR, respectively. AMR exhibits low resistance condition at $M \perp I_c$, i.e., $M//y$ or $M//z$ and high resistance condition at $M//x$. SMR, on the other hand, is magnetoresistance due to absorption or reflection of spin current generated in NM layer by SHE, giving the high resistance state at $M//z$ or $M//x$ and the low resistance state at $M//y$. Therefore, measuring the MR in the $xy$, $zx$, and $zy$ planes is essential to distinguish the contributions of SMR and AMR. It turns out that the angular dependence of $R$ in the $xy$ planes ($\theta_M = \pi/2$) should consist of $\Delta R_{AMR} + \Delta R_{SMR}$ while those in the $zx$ and $zy$ planes consist of only $\Delta R_{AMR}$ and $\Delta R_{SMR}$, respectively. Figure 2(d) shows the representative angular dependence of $R(\theta, \phi)$ for the CoFeB(3)/SIO(21) sample with $I_c$ and $B_{ext}$ being 1 mA and 1.35 T. Here, $\theta$ and $\phi$ represent the polar and azimuthal angle of $B_{ext}$. For the $xy$-plane, $B_{ext}$ is large enough to align $M$ along $B_{ext}$ yielding $\phi_M = \phi$. For the $zx$ and $zy$ planes, since $M$ is not aligned to $B_{ext}$ owing to the demagnetization effect, $\theta_M$ is characterized by $\theta_M = \cos^{-1}(R_H/R_{AHE})$, where $R_{AHE}$ is the saturated value of the anomalous Hall resistance and $R_H$ is Hall resistance concomitantly measured with SMR measurement at given $\theta$ of $B_{ext}$. We verified the saturation of $R_H$ within 0.7–1.1 T in all of the samples by performing Hall measurements, suggesting that the applied $B_{ext}$ of 1.35 T is sufficient to fully align the $M$ along the $z$ axis. The $R(\theta, \phi)$ curve is well reproduced by Eq. (2) for the $xy$, $zy$, and $zx$ planes (black solid lines in Fig. 2(d)), showing that the CoFeB/SIO bilayer exhibits a clear SMR signal in similarly to the previous studied CoFeB/NM bilayers[5-7,18]. Hereafter, we measure magnetoresistance in the $zy$ plane for characterizing $t_{FM}$ dependence of SMR.

Figure 3(a) shows $(R - R_0)/R_0$ in the $zy$ plane, the amplitude of which corresponds to the SMR ratio $\Delta R_{SMR}/R_0$, at $t_{CoFeB}$ = 3 and 6 nm, revealing an increasing trend with increasing $t_{CoFeB}$. Black solid lines represent the angular dependence corrected by $R_0$, following to the same procedure for Fig. 2(d). According to Eq. (1), the $\Delta R_{SMR}/R_0$ is influenced by $t_{CoFeB}$ through the three parameters: $x_{NM}$, $\xi_{SH}$, and $g_{FM}$. To elucidate the $t_{CoFeB}$ dependence of the $\Delta R_{SMR}/R_0$, we first compensated for the obvious $t_{CoFeB}$ dependence of $x_{NM}$ by dividing the absolute value of $\Delta R_{SMR}/R_0$ by $x_{NM}$. Figure 3(b) shows $t_{CoFeB}$ dependence of $|\Delta R_{SMR}/R_0|/x_{NM}$, exhibiting a clear linear relation in whole $t_{CoFeB}$ range up to 6 nm. In W/CoFeB (Ref. 9), as detailed in its supplemental information, $|\Delta R_{SMR}/R_0|/x_{NM}$ linearly increases up to $t_{CoFeB} \sim 1.5$ nm, which is below the spin diffusion length $\lambda_{CoFeB}$ of CoFeB, and then saturates in the range 1.5 nm < $t_{CoFeB}$ < 3 nm. This saturation behavior above $t_{CoFeB} \sim \lambda_{CoFeB}$ reflects the constant $\xi_{SH}$ value in the bulk limit of CoFeB. Our CoFeB/SIO bilayer still exhibits the linear dependence of $|\Delta R_{SMR}/R_0|/x_{NM}$ on $t_{CoFeB}$ in the range beyond $\lambda_{CoFeB}$, suggesting a significant contribution of $t_{FM}$-dependent $g_{FM}$ according to Eq. (1).

We now discuss the absorption of longitudinal spin current by the CoFeB layer. When $t_{NM}$ and $t_{FM}$ are sufficiently larger than $\lambda_{NM}$ = 1.4 nm for SIO[14] and $\lambda_{FM}$ = 1.0 nm for CoFeB[5], respectively, Eq. (1) is further transformed as follows.

$$\left|\frac{\Delta R_{SMR}}{R_0}\right|\frac{1}{x_{NM}} = \xi_{SH}^2\frac{\lambda_{NM}}{t_{NM}}\left[1-\frac{g_{FM}}{1+g_{FM}}\right], \qquad g_{FM} = \frac{(1-P^2)\rho_{NM}\lambda_{NM}}{\rho_{FM}\lambda_{FM}} \tag{3}$$

The second term, $-g_{FM}/(1+g_{FM})$, represents suppression of SMR by the contribution of longitudinal spin current absorption, which increases with $g_{FM}$. From a more microscopic perspective, $|\Delta R_{SMR}/R_0|/x_{NM}$ consequently increases with $P$, which is the only $t_{FM}$ dependent parameter in $g_{FM}$. This increasing trend of $|\Delta R_{SMR}/R_0|/x_{NM}$ with $P$ becomes more significant for the larger $\rho_{NM}\lambda_{NM}/\rho_{FM}\lambda_{FM}$. In Fig. 4(a), we plot $\rho_{NM}\lambda_{NM}/\rho_{FM}\lambda_{FM}$ as a function of $\rho_{NM}$ for SIO and widely-used metallic NMs such as Pt[8], W[5], and Ta[5]. SIO exhibits the largest $\rho_{NM}\lambda_{NM}/\rho_{FM}\lambda_{FM}$ because of its high $\rho_{NM}$ compared with the other NMs, suggesting that the contribution of longitudinal spin current absorption cannot be neglected in the CoFeB/SIO bilayers. This is based on the premise that the parameters of the NM layer ($\rho_{NM}$ and $\lambda_{NM}$) are not strongly influenced by $t_{CoFeB}$, and $\lambda_{FM}$ is a material constant. Finally, the constant value of $\rho_{FM}$ within the examined $t_{CoFeB}$ range has been experimentally validated in our previous work[18]. Consequently, our result that the SMR ratio increases with $t_{CoFeB}$ is ascribed to an enhancement of $P$ with increasing $t_{CoFeB}$, suggesting that the linear trend of the SMR ratio would still be observed as long as the thickness dependence of $P$ persists at CoFeB thickness above ~6 nm. While it is difficult to measure the $P$ at the current stage, the increasing trend of $P$ with thicker $t_{CoFeB}$ is qualitatively consistent with a previous report[36]; it discussed the thickness dependence of spin-transfer torque driven domain wall dynamics, which indicates that the torque becomes less effective at reduced thickness due to the critical role of $P$ in such dynamics.

In accordance with Eq. (3), we deduced the spin Hall angle $\xi_{SH}$ of SIO to be approximately 0.12 by adopting $P$ = 0.72 for a 2 nm-thick CoFeB film, which was estimated from spin-transfer torque driven magnetic domain wall dynamics[37)]. We emphasize the importance of using a spin polarization value at comparable film thicknesses used in our study (below 6 nm), rather than a value obtained from a bulk-like 200 nm-thick CoFeB film measured by a point-contact Andreev reflection technique[38)]. The obtained $\xi_{SH}$ in the CoFeB/SIO bilayers is comparable to $\xi_{SH} \sim 0.10$ for Pt[4,9,12,14,18,23)] but somewhat smaller than values estimated from other techniques, such as the harmonic Hall method (~0.30) in our previous study[18)] and spin-torque ferromagnetic resonance (~0.20–0.50) reported by other groups[14,15,21)]. Although the precise reason for this variation remains to be clarified, the obtained $\xi_{SH}$ of 0.12 is still within a reasonable range for efficient spin current generation from NMs. It is also worth noting that this value is significantly higher than $\xi_{SH}$ = 0.001–0.005 reported for CoFeB single layers[39,40)], indicating that spin current generation stemming from the SHE in FM layer[39 – 41)] is negligibly small in our system.

To further clarify the effect of $g_{FM}$ across bilayers with different NMs, the absolute value $|\xi_{SH}|$ as a function of $\rho_{NM}\lambda_{NM}/\rho_{FM}\lambda_{FM}$ is summarized in Fig. 4(b). The $\xi_{SH}$ was obtained via Eq. (3) with the $g_{FM}$ term excluded [$g_{FM}$ = 0] or via Eq. (3) with the $g_{FM}$ term included [$g_{FM} \neq 0$], with 2 nm-thick CoFeB and $P$ = 0.72 for all the CoFeB/NM bilayers. While all the $|\xi_{SH}|$ values at $g_{FM} \neq 0$ are higher than those at $g_{FM}$ = 0, the discrepancy in $|\xi_{SH}|$ becomes more pronounced for larger $\rho_{NM}\lambda_{NM}/\rho_{FM}\lambda_{FM}$ ratios. Figure 4(c) shows the relative correction of $|\xi_{SH}|$ for the bilayers, providing ~5% for CoFeB/Pt, ~20% for CoFeB/Ta, ~19% for CoFeB/W, and ~71% for CoFeB/SIO in order of $\rho_{NM}\lambda_{NM}/\rho_{FM}\lambda_{FM}$ ratio. Here, the relative correction is defined as $\left(\frac{|\xi_{SH}|[g_{FM} \neq 0]}{|\xi_{SH}|[g_{FM} = 0]}-1\right) \times 100$ (%). This result highlights that the $|\xi_{SH}|$ value of 0.07 at $g_{FM}$ = 0 for our CoFeB/SIO is a largely underestimated, reinforcing necessity of considering $g_{FM}$. We discuss the influence of thickness on the relative correction. Considering a standard point of $P$ = 0.72 at 2 nm in Fig. 4(c), thicker CoFeB, which is expected to exhibit a higher $P$ value, reduces the $g_{FM}$ term, leading to a decrease of the relative correction due to suppression of the underestimation of the spin Hall angle. Thus, the high resistivity of SIO plays a critical role in the absorption of longitudinal spin current by CoFeB, emphasizing the significant influence of the magnetic layer on the SMR ratio. Moreover, because resistive materials used as spin current sources are expected to be an asset for highly sensitive sensors based on spin current detectors, the present study clarifying the spin transport using SIO via SMR represents an important step toward practical applications.

## 4. Conclusion

In conclusion, we investigated the magnetic layer thickness dependence of SMR in CoFeB/SIO bilayers. The SMR ratio is found to increase with increasing CoFeB layer thickness, which is qualitatively consistent with an SMR model taking into account the absorption of longitudinal spin current by the magnetic layer. The effective spin Hall angle is corrected from the initial value of 0.07 to 0.12, a ~71% increase via the SMR model, highlighting that the absorption is particularly significant in our bilayer system utilizing SIO as a highly-resistive NM material, instead of commonly used metallic NMs such as Pt. Our result indicates that the magnetic layer has a significant effect on the SMR ratio in bilayers with highly-resistive NM materials. The present study provides a clue toward more practical applications for highly sensitive sensors based on spin current detectors, which can be driven by highly-resistive NM materials including topological and two-dimensional materials.

## ACKNOWLEDGEMENT

The authors thank T. Arakawa for technical support. This work was supported by Nanotechnology Platform of MEXT, the JSPS KAKENHI (Grant Nos. JP19K15434, JP19H05823, 22H04478, and 24H01666), JPMJCR1901 (JST-CREST), the MEXT "Spintronics Research Network of Japan (Spin-RNJ)", and Nippon Sheet Glass foundation for Materials Science and Engineering, and Iketani Science and Technology Foundation. We acknowledge the stimulating discussions at the meeting of the Cooperative Research Project of the Research Institute of Electrical Communication, Tohoku University.

## Appendix A. Surface morphology and electrical transport property of $SrIrO_3$ film

Figure A1(a) displays an AFM image of 20 nm-thick SIO film grown on $DyScO_3$ substrate, exhibiting an RMS roughness of 0.2 nm. This atomically flat surface is expected to reduce potential interface scattering effects in the bilayer, which could otherwise affect the SMR measurement. To characterize the electrical transport property of the SIO film, temperature dependence of resistivity was measured using the standard four-terminal method as shown in Fig. A1(b). The resistivity shows only weak temperature dependence, indicative of the semimetallic behavior typical of $SrIrO_3$, consistent with previously reported PLD-grown films[21,42–44]. It has been reported that Ir cation deficiency in SIO, often caused by low oxygen partial pressure during deposition, significantly increases the resistivity, often exceeding 2 mΩcm at room temperature[21]. In contrast, the relatively low resistivity observed in our film suggests that Ir vacancies are negligible and the film is of high quality.

## AUTHOR DECLARATIONS

Conflict of Interest

The authors have no conflicts to disclose.

## AUTHOR CONTRIBUTIONS

S. Hori: Data Curation (lead); Formal Analysis (lead); Visualization (equal); Writing/Original Draft Preparation (equal). K. Ueda: Conceptualization (lead); Visualization (equal); Writing/Original Draft Preparation (equal). J. Shiogai: Writing/Review & Editing (equal). J. Matsuno: Supervision (lead); Writing/Review & Editing (equal).

## DATA AVAILABILITY

The data that support the findings of this study are available from the corresponding authors upon reasonable request.

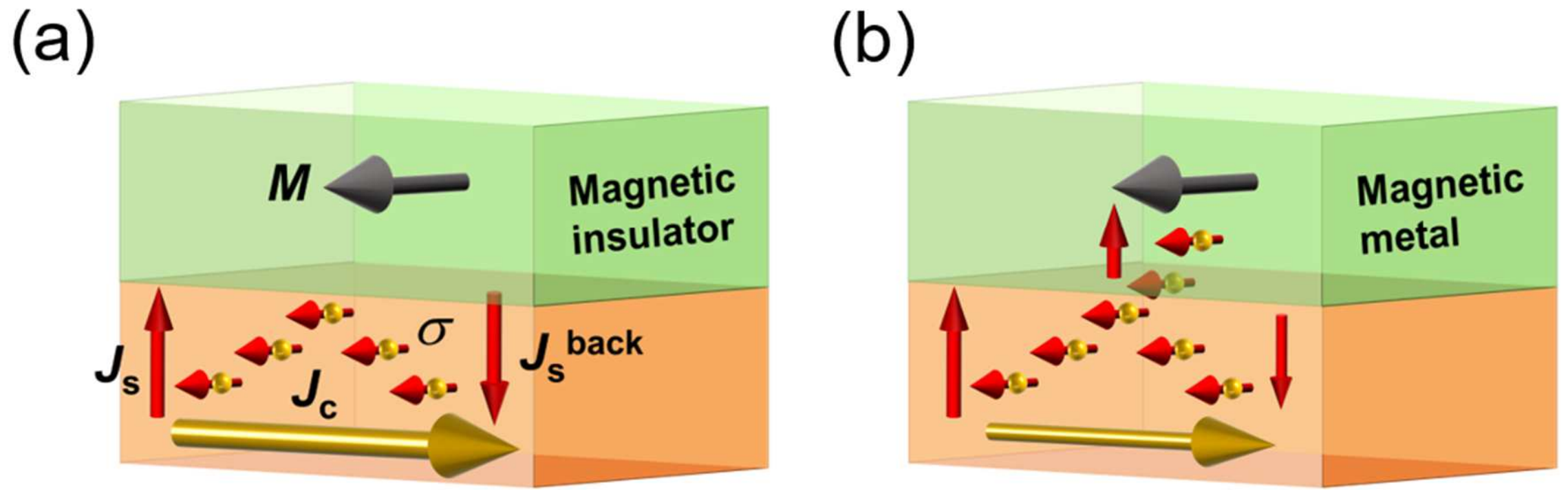

Fig. 1 Schematic illustration of the spin current ($J_s$) absorption by magnetic layer in the case of $M$ // $\sigma$. (a) Magnetic insulator: Longitudinal spin current is not absorbed by the magnetic layer, resulting in spin back flow ($J_s^{back}$). (b) Magnetic metal: Longitudinal spin current is absorbed by the magnetic layer, leading to reduced spin back flow into the NM layer.

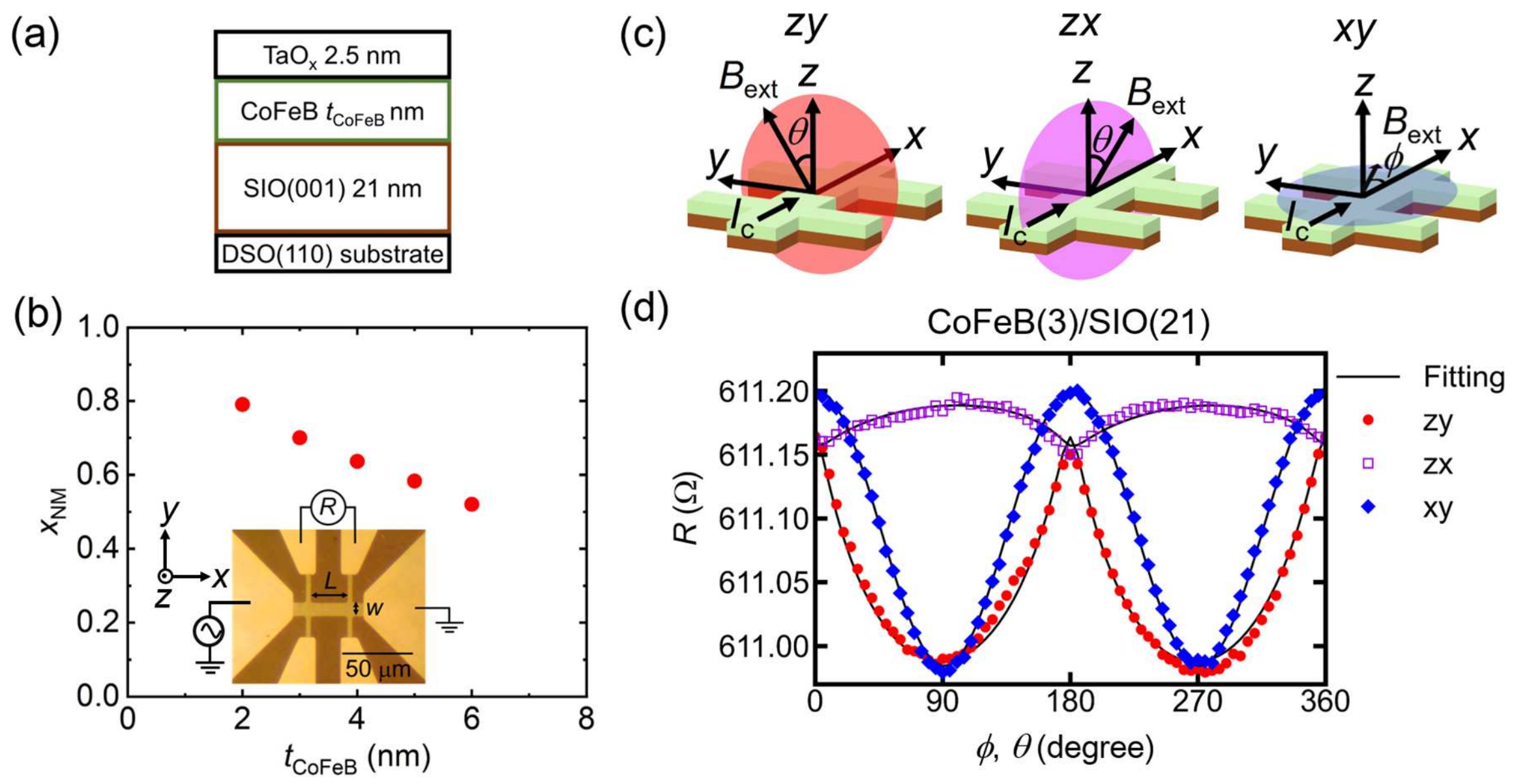

Fig. 2 (a) Schematic of the layer structures. (b) Current fraction flowing through non-magnet $x_{NM}$ as a function of CoFeB thickness $t_{CoFeB}$. (Inset) Optical microscopy image of a Hall-bar device and the measurement configuration. (c) Illustration of the rotation planes with applied magnetic field ($B_{ext}$ = 1.35 T). Charge current ($I_c$) is applied along the *x*-axis direction. (d) Angular dependence of longitudinal resistance for SIO sample in *zy* (red circles), *zx* (open purple squares), and *xy* (blue diamonds) planes. Solid black curves are fitting curves to the data using Eq. (2).

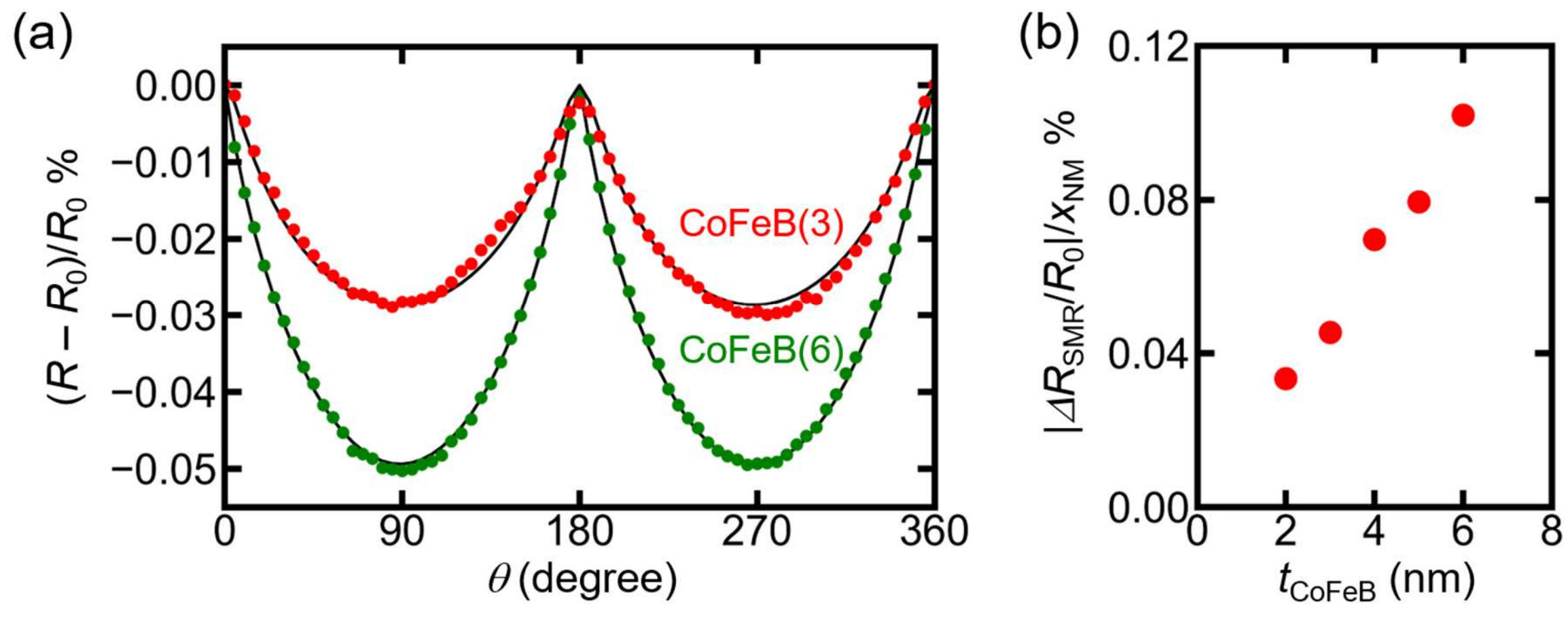

Fig. 3 (a) Magnetic field angular dependence $(R - R_0)/R_0$ for $t_{CoFeB}$ = 3 (red circles) and 6 nm (green diamonds) in *zy* plane. Black solid lines represent fits to the angular dependence according to the procedure described in main text. (b) The $t_{CoFeB}$ dependence of $\Delta R_{SMR}/R_0$ normalized by current fraction flowing through SIO ($x_{NM}$).

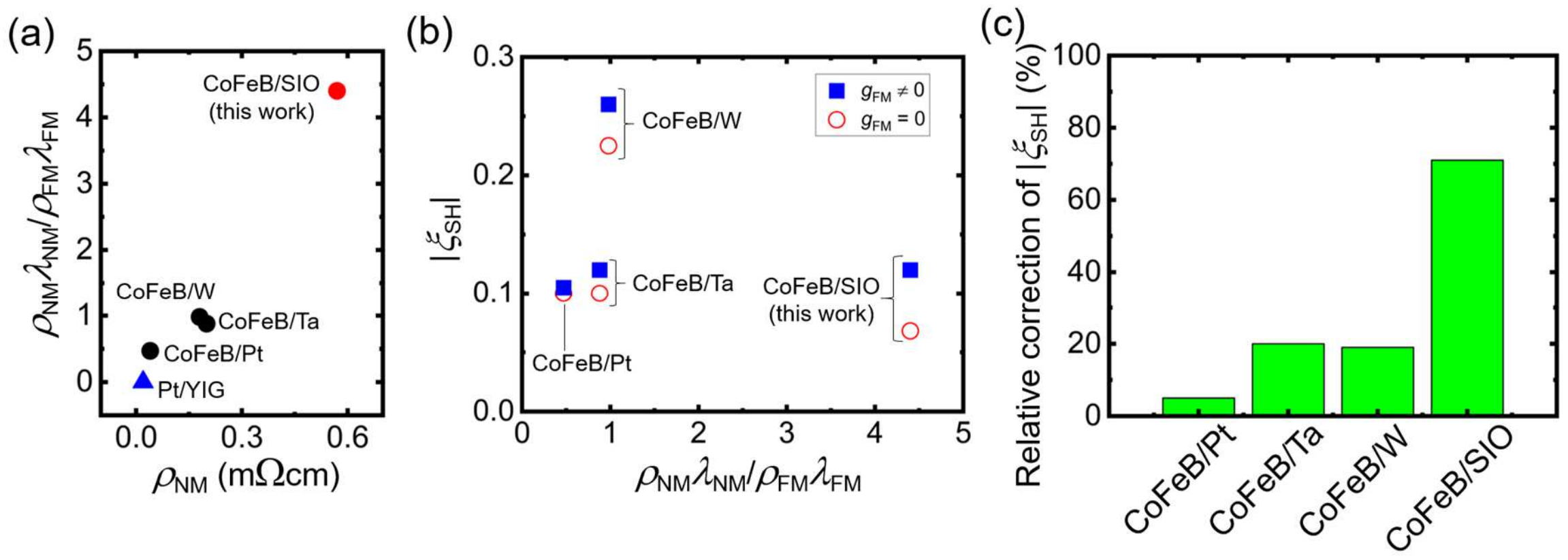

Fig. 4 (a) The $\rho_{NM}\lambda_{NM}/\rho_{FM}\lambda_{FM}$ ratios are plotted as a function of the resistivity of the non-magnetic layer $\rho_{NM}$ for CoFeB/SIO (this work), Pt/YIG[1)], CoFeB/Pt[8)], CoFeB/W[5)], and CoFeB/Ta[5)] bilayers. (b) The absolute value of effective spin Hall angle ($\xi_{SH}$) as a function of the $\rho_{NM}\lambda_{NM}/\rho_{FM}\lambda_{FM}$ ratio for CoFeB/(Pt, Ta, W, SIO) bilayers, with/without $g_{FM}$ term. (c) Relative correction of $|\xi_{SH}|$ versus bilayers based on (b).

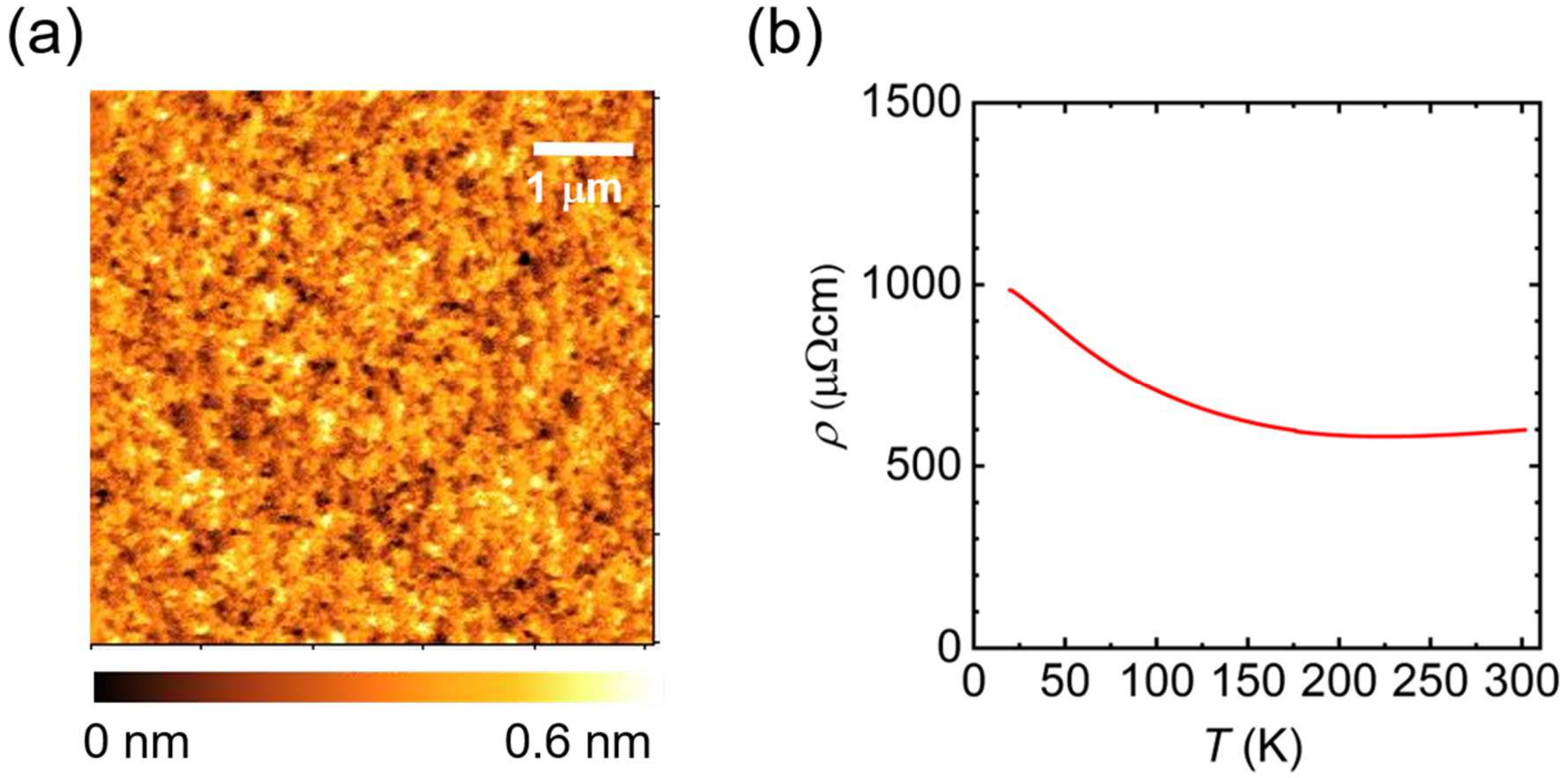

Fig. A1 (a) Atomic force microcopy image and (b) temperature dependence of resistivity of 20 nm-thick-SIO film grown on DSO(110) substrate.